\documentclass{article}

\usepackage{spconf,amsmath,theorem,epsfig,amssymb}
\usepackage{amsfonts}
\usepackage{pst-all}

\newcommand{\F}{\mathbb{F}}
\newtheorem{thm}{Theorem}

\title{Efficient FPGA-based multipliers for $\F_{3^{97}}$ and $\F_{3^{6 \cdot 97}}$}

\threeauthors
  {Jamshid Shokrollahi
	\sthanks{Partially funded by the German Research Foundation (DFG) under
project RU 477/8}}
  {B-IT\\ University of Bonn \\ Germany \\
   jamshid@bit.uni-bonn.de}
{Elisa Gorla
  \sthanks{Partially funded by the Swiss National Science Foundation under
grant no. 107887}}
  {Department of Mathematics \\ University of Z\"urich\\ Switzerland \\
   elisa.gorla@math.unizh.ch}
  {\setcounter{footnote}{1}Christoph Puttmann$^{\fnsymbol{footnote}}$}
  {Heinz Nixdorf Institute\\ University of Paderborn \\ Germany \\
   puttmann@hni.upb.de}

\begin{document}

\maketitle

\begin{abstract}
In this work we present a new structure for multiplication in finite fields. 
This structure is based on a digit-level LFSR (Linear Feedback Shift
Register) multiplier in which the area of digit-multipliers are
reduced using the Karatsuba method. We compare our results with the
other works of the literature for $\F_{3^{97}}$. We also propose new
formulas for multiplication in $\F_{3^{6 \cdot 97}}$. These new
formulas reduce the number of $\F_{3^{97}}$-multiplications from $18$
to $15$. The finite fields $\F_{3^{97}}$ and $\F_{3^{6 \cdot 97}}$ are
important fields for pairing based cryptography. 
\end{abstract}


\section{Introduction}
\label{sec:intro}

Efficient multiplication in finite fields is a central task in the
implementation of most public key cryptosystems. A great amount of
work has been devoted to this topic (see \cite{knu98} or
\cite{gatger03} for a comprehensive list).
The two types of finite fields which are mostly used in cryptographic
standards are binary finite fields of type $\F_{2^{m}}$ and prime
fields of type $\F_{p}$, where $p$ is a prime (cf. \cite{dss00}).
Efforts to efficiently fit finite field arithmetic into commercial
processors resulted into applications of medium characteristic finite
fields like those reported in \cite{baipaa98} and \cite{avamih03}.
Medium characteristic finite fields are fields of type $\F_{p^{m}}$,
where $p$ is a prime slightly smaller than the word size of the
processor, and has a special form that simplifies the modular
reduction. Mersenne prime numbers constitute an example of primes
which are used in this context. 
The security parameter is given by the length of the binary
representations of the field elements, and the extension degree $m$ is
selected appropriately.
Due to security considerations, the extension degree for fields of
characteristic $2$ or medium characteristic is usually chosen to be
prime.

With the introduction of the method of Duursma and Lee for the
computation of the Tate pairing (cf. \cite{duulee03}), fields of type
$\F_{3^{m}}$ for $m$ prime have attracted special attention.
Computing the Tate pairing on elliptic curves defined over
$\F_{3^{m}}$ requires computations both in $\F_{3^m}$ and in
$\F_{3^{6m}}$.
In \cite{kermar05} calculations are implemented using the tower of
extensions $$\F_{3^m}\subset\F_{3^{2m}}\subset\F_{3^{6m}}$$ and the
inherent parallelism of multiplication in extension fields is used to
accelerate the operations. 
Hardware designs and especially FPGA-based ones are suitable platforms
for parallel implementation of algorithms.
In that work multiplications in the first and the second field
extensions are computed via $3$ and $6$ multiplications in the ground
fields, respectively, requiring $18$ multiplications in $\F_{3^{97}}$.

In our current work, which is mostly based on \cite{kermar05}, on the
one hand, we use asymptotically fast methods to improve the
performance of multiplication in $\F_{3^{97}}$, and on the other hand,
we propose new multiplication formulas to speedup multiplication in
$\F_{3^{6 \cdot 97}}$.
Using the new formulas, multiplication in $\F_{3^{6 \cdot 97}}$ is
done with only $15$ multiplications instead of $18$. 
We use the same extension tower, using $3$ multiplications in
$\F_{3^{97}}$ to multiply elements in $\F_{3^{2 \cdot 97}}$, but only
$5$ multiplications in $\F_{3^{2 \cdot 97}}$ for $\F_{3^{6 \cdot
    97}}$.
Our proposed method has a slightly increased number of
additions in comparison to the Karatsuba method. Notice however that a
multiplication in $\F_{3^{97}}$ requires many more resources than an
addition, therefore the overall resource consumption will be reduced.
The details of our method to generate the new formulas have been
omitted to limit the complexity and diversity of materials in this
paper, and have been submitted as another paper for CHES 2007.

A consistent amount of work has been done on hardware-based
multiplication in finite fields, especially those of characteristic
$3$. The authors of \cite{bergua03} propose a least significant
digit-element (LSDE) multiplier for $\F_{3^{m}}$. 
This multiplier divides the input polynomials into digits of length D. Whereas the digits of one input polynomial are processed in parallel, the digits of the other input polynomial are handled serially. Then the result is reduced modulo the
irreducible polynomial. 
The same structure has also been used in \cite{kermar05} for
multiplication in $\F_{3^{97}}$. 
Our multiplier, on the other hand, is based on the digit-serial
implementation of LFSR (Linear Feedback Shift Register) multiplier
which is widely used in the literature (see \cite{mce87} or
\cite{sho06}), and performs the modular reduction during the multiplication. The first contribution of our current work is
the application of the Karatsuba multiplier inside the
digit-multipliers, which results in smaller area for these
multipliers. 
Our results demonstrate the efficiency of this design compared to other works. 
The second contribution is the application of a method using only $5$
multiplications in $\F_{3^{2 \cdot 97}}$ for multiplication in
$\F_{3^{6 \cdot 97}}$. This results in an area-saving of almost
$17\%$ compared to the Karatsuba method which is used in
\cite{kermar05}.
   
Our work is organized as follows. Section~\ref{sec:multf397} is
devoted to the general structure of our multiplier for $\F_{3^{97}}$. In
Section~\ref{sec:circuit} we describe some improvements on the
traditional LFSR multiplier and compare our results with other works
from the literature. 
In Section~\ref{sec:multf3697} the new formulas for $\F_{3^{6 \cdot
    97}}$ together with suggestions for a new multiplier are
presented, and Section~\ref{sec:conclusion} concludes the paper.

\section{Multiplication in $\F_{3^{97}}$}
\label{sec:multf397}

The finite field $\F_{3^{97}}$ can be represented as a vector space
over $\F_{3}$. In this representation, elements of $\F_{3^{97}}$ are
vectors of length $97$ over $\F_{3}$. Addition of elements is computed
by adding corresponding vectors. Multiplication is more complicated,
and depends on the selected basis for $\F_{3^{97}}$. There are two
popular bases which are used often in the literature, namely
polynomial and normal bases. 
A polynomial basis is generally more suitable for multiplication,
hence we choose this basis in our work. 

In the polynomial basis, elements of $\F_{3^{97}}$ are represented as
polynomials of degree at most $96$ over $\F_3$. Two elements are added
by adding of the corresponding polynomials. Multiplication is based on polynomial
multiplication followed by reduction modulo the irreducible polynomial,
which generates the polynomial basis. In our case the irreducible
polynomial, which we denote by $f(x)$, is 
\begin{equation}
x^{97}+x^{16}+2.
\label{equ:irreducible}
\end{equation}
In the next sections we show the details of polynomial arithmetic in
our designs.

\subsection{Arithmetic in $\F_{3}$}

The element $a \in \F_{3}$ is represented using the vector $(a_1,
a_0)$ of two bits such that the elements $0$, $1$, and $2$ are
$(0,0)$, $(0,1)$, $(1,0)$, respectively. In this representation the
operations addition, multiplication, and negation (multiplication by
$2$) are done, as shown in \cite{grapag05}, using
Equations~\ref{equ:bitadd}, \ref{equ:bitmult}, and \ref{equ:bitneg},
respectively. 
\begin{align}
\label{equ:bitadd} & (a_1, a_0)+(b_1,b_0)  = ((a_0 \vee b_0) \oplus t, (a_1 \vee b_1) \oplus t),\\
\nonumber & \text{where } t= (a_0 \vee b_1) \oplus (a_1 \vee b_0)\\
\label{equ:bitmult}  & (a_1, a_0) \cdot (b_1,b_0) = ((a_1 \wedge b_0) \vee (a_0 \wedge b_1),\\
\nonumber &  (a_0 \wedge b_0) \vee (a_1 \wedge b_1),\\
\label{equ:bitneg} & -(a_1, a_0) = (a_0, a_1).  
\end{align}
The implementation of Equations~\ref{equ:bitadd} and \ref{equ:bitmult}
is done using 2 LUTs in the FPGA, whereas (\ref{equ:bitneg}) is only a
permutation of bits.

\subsection{Structure of the multiplier for $\F_{3^{97}}$}

\begin{figure}
\includegraphics[width=8cm]{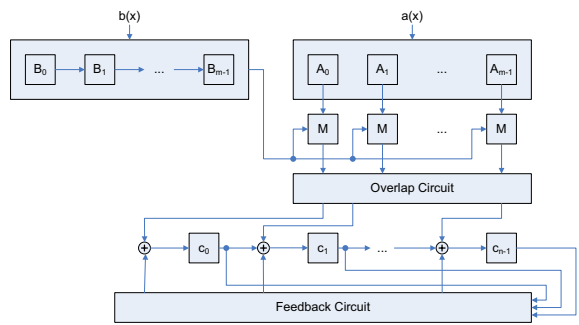}
\caption{Structure of a digit-level LFSR multiplier}
\label{fig:lfsr}
\end{figure}

The structure of a digit-level LFSR multiplier is shown in
Figure~\ref{fig:lfsr}. In this figure the two input polynomials
$a(x)$, and $b(x)$ are loaded into registers $A$ and $B$,
respectively, and divided into digits of length $D$. In each clock
cycle the most significant digit of $B$ is multiplied by the words of
$A$, through digit-multipliers specified by M, and added to the
content of the register in the feedback circuit.
Inputs to the digit multipliers are two polynomials of degree $D-1$ in $x$.
The product is a polynomial of degree $2(D-1)$. Powers $x^D$ to
$x^{2(D-1)}$ of each multiplier must be added to the powers $x^0$ to
$x^{D-2}$ of the next multiplier. This is done by the overlap circuit.
In each clock cycle the register $B$ and LFSR are shifted by $D$ bits to the right.
Shifting LFSR to right is equivalent to multiplication by $x^D$ which
generates the powers $x^{97}$ to $x^{96+D}$. These powers are reduced
modulo $f(x)$ of (\ref{equ:irreducible}) using the feedback circuit. 
The name Linear Feedback Shift Register descends from these feedback structures. 
 For more information about the digit-level LFSR multiplier and its
 costs for classical methods see \cite{sho06}. In the next section we
 discuss our improvements to the traditional LFSR multiplier.  

\section{The Karatsuba method}
\label{sec:circuit}

In this section we use asymptotically fast methods to reduce the size of digit-multipliers. 
We use a similar approach to \cite{gatsho06} and combine the classical
and the Karatsuba methods to build small digit-multipliers.
Two linear polynomials $a_1x+a_0$ and $b_1x+b_0$ are multiplied
classically using the formula
\begin{equation}
a_1b_1x^2+(a_1b_0+a_0b_1)x+a_0b_0
\label{equ:classical}
\end{equation}
with $4$ multiplications and $1$ addition. 
The same product can also be computed via
\begin{equation}
a_1b_1x^2+((a_1+a_0)(b_1+b_0)-a_1b_1-a_0b_0)x+a_0b_0.
\label{equ:karatsuba}
\end{equation}
The new formula is called the Karatsuba method (see
\cite{karofm63}). It requires $7$ operations instead of $5$, but only
$3$ multiplications, and uses fewer resources when the coefficients
$a_0,a_1,$ $b_0,b_1$ are replaced by polynomials. 
The classical method for multiplication of two polynomials of degree
$n-1$ requires $O(n^2)$ operations. Recursive application of
the Karatsuba method reduces the cost of a multiplication to
$O(n^{1.59})$ operations.
We represent the classical multiplication of two polynomials of degree
$n-1$ by $\mathcal{C}_n$ and the method of (\ref{equ:karatsuba}) by
$\mathcal{K}$.
The methods $\mathcal{C}_n$ for $n \in \mathbb{N}$, and $\mathcal{K}$
constitute a set of polynomial multiplication methods. We call this set
$\sf{T}$. Using the elements of $\sf{T}$ we define the set of
recursive multiplication methods $\sf{T}^{*}$ which contains the
elements of $\sf{T}$ and all recursive combinations of elements of
$\sf{T}^{*}$. The recursive combination of the two methods
$\mathcal{M}$ and $\mathcal{N}$, for polynomials of lengths $m$ and
$n$, respectively, is the multiplication method
$\mathcal{M}\mathcal{N}$ for polynomials of length $mn$. Let
\begin{align*}
& a(x) = a_{mn-1}x^{mn-1}+\cdots+a_0, \text{ and}\\
 & b(x) = b_{mn-1}x^{mn-1}+\cdots+b_0
\end{align*}
be given polynomials. In order to apply $\mathcal{M}\mathcal{N}$, we
write these polynomials as 
\begin{align*}
& a(x) = A_{m-1}X^{m-1}+\cdots+A_{0}, \text{ and}\\
 & b(x) = B_{m-1}X^{m-1}+\cdots+B_{0},
\end{align*}
where $X=x^n$ and $A_0, \cdots A_{m-1}, B_0, \cdots B_{m-1}$ are
polynomials of degree $n-1$.
If the polynomials $A_i$ and $B_i$ were coefficients,
the two polynomials $a(x)$ and $b(x)$ would be multiplied using
$\mathcal{M}$. The product using the method $\mathcal{M}\mathcal{N}$
consists of several multiplications of the polynomials $A_i$ and
$B_i$, which are performed using $\mathcal{N}$.
We implement the digit-multipliers using the elements of $\sf{T}^{*}$
to reduce their size. Our approach is similar to \cite{gatsho06}.

In Table~\ref{tab:lfsr397costs} we show the results of implementing
$\F_{3^{97}}$ multipliers on a XC2VP20-6FF896 FPGA. In this
table the first column is the digit-size $D$. In a digit-level
multiplier with digit-size $D$, inputs are preceded by enough zeros
so that their length becomes a multiple of $D$. Hence it is natural to
choose a value of $D$ such that the difference $\lceil m / D \rceil -
m/D$ is as small as possible. Our values for $D$ are selected using
this criteria and hence differ from other standard values like
multiples of $4$ in other works (see \cite{bergua03} and
\cite{kermar05}). 
The string in the second column shows the recursive combination of the
Karatsuba and classical methods which is applied.
It is important to notice that the method $\mathcal{K}\mathcal{C}_2$,
which we used for polynomials of degree $6$, applies to polynomials of
length $7$. 
Therefore, we add a zero in front of the polynomial and then remove all
the gates containing an operation with the coefficients which are
known to be zero.
Hence this multiplier requires fewer resources than a complete
$\mathcal{K}\mathcal{C}_2$. This point distinguishes our
approach from that in \cite{gatsho06}. In the third, fourth, and fifth
columns are the number of slices, maximum working frequency of the
multiplier, and the required clock cycles for our designs.   

\begin{table}
\caption{Timing and area costs of digit-level LFSR multipliers in
  $\F_{3^{97}}$ for different values of digit-size $D$}
\begin{scriptsize}
\begin{tabular}{|c|c|c|c|c|}\hline
{$D$} & {\bf Multiplication} & {\bf \# of slices} & {\bf Maximum} & {\bf \# of clock}\\
 & & & {\bf frequency (MHz)}& {\bf cycles} = $\lceil 97/D \rceil$\\ \hline
$1$ & $-$ & $327$ & $300$ & $97$\\
$2$ & $\mathcal{C}_{2}$ & $800$ & $174$ & $49$\\
$4$ & $\mathcal{C}_{4}$ & $1716$ & $125$ & $25$\\
$7$ & $\mathcal{K}\mathcal{C}_{4}$ & $2954$ & $111$ & $14$\\
$14$ & $\mathcal{K}\mathcal{K}\mathcal{C}_{4}$ & $4006$ & $72$ & $7$\\ \hline
\end{tabular}
\end{scriptsize}
\label{tab:lfsr397costs}
\end{table}
 
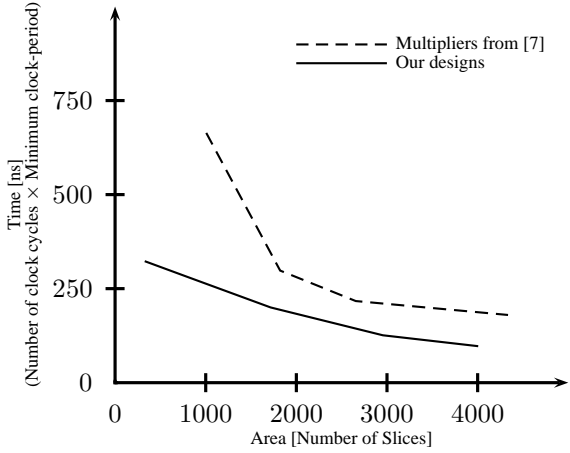
\begin{figure}

  \def\dataKerMar05{%
		1006 664 1821 298 2655 217 4335 180
    }
  \def\dataOurs{%
		327 323 800 281 1716 200 2954 126 4006 97
  }
\psset{xunit=0.0012cm,yunit=0.005cm}
  \centering
\begin{pspicture}(5000,-200)(-1200,1200)
  \psaxes[Dx=1000,Dy=250,linewidth=1.2pt]{->}(0,0)(5000,1000)
  \listplot{\dataOurs}
  \listplot[linestyle=dashed]{\dataKerMar05}
  \rput(2500,-150){\scriptsize Area [Number of Slices]}
  \rput{90}(-1100,500){\scriptsize Time [ns]}
  \rput{90}(-900,500){\scriptsize (Number of clock cycles $\times$
  Minimum clock-period)}
  \rput[l](2000,850){\psline(0,0)(1000,0)}
  \rput[l](3100,850){\scriptsize Our designs}
  \rput[l](2000,900){\psline[linestyle=dashed](0,0)(1000,0)}
  \rput[l](3100,900){\scriptsize Multipliers from \cite{kermar05}}
\end{pspicture}\hspace{0.75em}

\caption{Time vs. area comparisons of our multipliers with those in \cite{kermar05}}
\label{fig:lfsrcomp}
\end{figure}

The results of comparing our results with those in \cite{kermar05} are
shown in Figure~\ref{fig:lfsrcomp}. Different digit-levels result in
different circuits, which we compare with respect to both time and
area. Area is the number of slices, whereas time is the product of
clock cycles and minimum period. Both designs are on the same
technology, but the speed grade of the FPGA in \cite{kermar05} is not
available.
As it is shown, our designs have better area-time performance. 
These improvements result, on the one hand, by using asymptotically
faster methods, and on the other hand, by integrating the modular
reduction stage into the LFSR. When a small digit-serial multiplier is
used even the small size of a modular reduction must be taken into account.

\section{Multiplication in $\F_{3^{6 \cdot 97}}$}
\label{sec:multf3697}

Multiplication in $\F_{3^{6 \cdot 97}}$ is done in the same way as in
\cite{kermar05}, as a tower of extensions of degrees $2$ and $3$, i.e.
$$
\begin{array}{lcl}
\F_{3^{97}} & \cong & \F_{3}/(x^{97}+x^{16}+2)\\ 
\F_{3^{2 \cdot 97}} & \cong & \F_{3^{97}} / (y^2+1)\\
\F_{3^{6 \cdot 97}} & \cong & \F_{3^{2 \cdot 97}} / (z^{3}-z-1).\\
\end{array}
$$
The elements of $\F_{3^{2 \cdot 97}}$ are polynomials of degree $1$ in
$s$ over $\F_{3^{97}}$, for $s$ a root of $y^2+1$ in $\F_{3^{2 \cdot 97}}$.
The polynomials are multiplied by applying (\ref{equ:karatsuba}) and then
reduced modulo $s^2+1$. The elements of $\F_{3^{6 \cdot 97}}$ are
polynomials of degree $3$ in $r$, a root of $z^3-z-1$ in $\F_{3^{6
    \cdot 97}}$. They are multiplied using the formulas
(\ref{equ:poly4formulas}) and then reduced modulo $r^3-r-1$.
\begin{equation}
\begin{array}{l}
(a_0+a_1r+a_2r^2)(b_0+b_1r+b_2r^2)=\\
c_0+c_1r+c_2r^2+c_3r^3+c_4r^4, \text{where}\\
P_0 = (a_{0}+a_{1}+a_{2})(b_{0}+b_{1}+b_{2})\\
P_1 = (a_{0}+sa_{1}-a_{2})(b_{0}+sb_{1}-b_{2})\\
P_2 = (a_{0}-a_{1}+a_{2})(b_{0}-b_{1}+b_{2})\\
P_3 = (a_{0}-sa_{1}-a_{2})(b_{0}-sb_{1}-b_{2})\\
P_4 = a_2b_2, \text{ and}\\
c_{0}=P_{0}+P_{1}+P_{2}+P_{3}-P_{4}\\
c_{1}=P_{0}-sP_{1}-P_{2}+sP_{3}\\
c_{2}=P_{0}-P_{1}+P_{2}-P_{3}\\
c_{3}=P_{0}+sP_{1}-P_{2}-sP_{3}\\
c_{4}=P_{4},
\end{array}
\label{equ:poly4formulas}
\end{equation}
Combining (\ref{equ:karatsuba}), (\ref{equ:poly4formulas}) we have the
following theorem.

\begin{thm}
Let $\alpha, \beta \in \F_{3^{6 \cdot 97}}$ be given as:
\begin{align*}
\alpha = & a_0+a_1s+a_2r+a_3rs+a_4r^2+a_5r^2s\\
\beta = & b_0+b_1s+b_2r+b_3rs+b_4r^2+b_5r^2s.\\
\end{align*}
Let further their product $\gamma = \alpha \beta \in \F_{3^{6 \cdot 97}}$ be 
\begin{align*}
\gamma = c_0+c_1s+c_2r+c_3rs+c_4r^2+c_5r^2s.\\
\end{align*}
Then the coefficients $c_0 \cdots c_5$ of the product can be computed
using only $15$ multiplications in $\F_{3^{97}}$.
\end{thm}

Closed-form formulas for this multiplication are shown in Appendix
\ref{sec:appendix}. Scalar multiplications are particularly simple
using these formulas. Scalar multiplications are multiplications by
$-1$, $s$, and $-s$. Negation of coefficients and consequently of
polynomials is only a permutation of bits, as seen in
Section~\ref{sec:multf397}.
Indeed multiplication of an element in $\F_{3^{2 \cdot 97}}$ by $s$ is
a permutation, too. 
Let $\alpha = a_1s+a_0 \in \F_{3^{2 \cdot 97}}$, then
$$s \alpha = a_1s^2+a_0s \mod s^2+1 = a_0s-a_1.$$ 

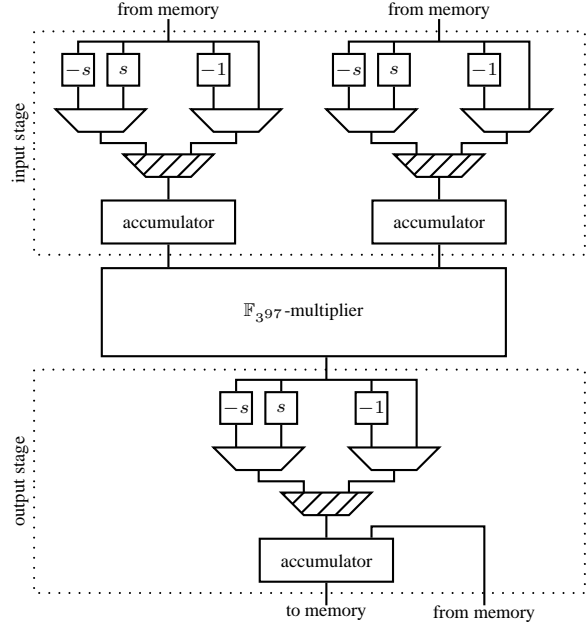
\begin{figure}
\centering
\psset{unit=0.3cm}
\begin{pspicture}(-2,-2)(24,28)
\rput[b](5,25){\scriptsize from memory}
\rput[b](17,25){\scriptsize from memory}
\rput[lb](0,15){
	\psframe(2,0)(8,2)
	\rput(5,1){\scriptsize accumulator}
	\psline(5,2)(5,3)
	\pspolygon[fillstyle=hlines](4,3)(6,3)(7,4)(3,4)
	\pspolygon(1,5)(3,5)(4,6)(0,6)
	\psline(2,5)(2,4.5)(4,4.5)(4,4)
	\pspolygon(7,5)(9,5)(10,6)(6,6)
	\psline(8,5)(8,4.5)(6,4.5)(6,4)
	\psframe(0.25, 7)(1.75, 8.5)
	\rput(1,7.75){\scriptsize $-s$}
	\psline(1,7)(1,6)
	\psframe(2.25, 7)(3.75, 8.5)
	\rput(3,7.75){\scriptsize $s$}
	\psline(3,7)(3,6)
	\psframe(6.25,7)(7.75,8.5)
	\rput(7,7.75){\scriptsize $-1$}
	\psline(7,7)(7,6)
	\psline(9,6)(9,9)(1,9)(1,8.5)
	\psline(3,8.5)(3,9)
	\psline(7,8.5)(7,9)
	\psline(5,9)(5,10)
}
\rput[lb](12,15){
	\psframe(2,0)(8,2)
	\rput(5,1){\scriptsize accumulator}
	\psline(5,2)(5,3)
	\pspolygon[fillstyle=hlines](4,3)(6,3)(7,4)(3,4)
	\pspolygon(1,5)(3,5)(4,6)(0,6)
	\psline(2,5)(2,4.5)(4,4.5)(4,4)
	\pspolygon(7,5)(9,5)(10,6)(6,6)
	\psline(8,5)(8,4.5)(6,4.5)(6,4)
	\psframe(0.25, 7)(1.75, 8.5)
	\rput(1,7.75){\scriptsize $-s$}
	\psline(1,7)(1,6)
	\psframe(2.25, 7)(3.75, 8.5)
	\rput(3,7.75){\scriptsize $s$}
	\psline(3,7)(3,6)
	\psframe(6.25,7)(7.75,8.5)
	\rput(7,7.75){\scriptsize $-1$}
	\psline(7,7)(7,6)
	\psline(9,6)(9,9)(1,9)(1,8.5)
	\psline(3,8.5)(3,9)
	\psline(7,8.5)(7,9)
	\psline(5,9)(5,10)
}
\rput[lb](7,0){
	\psframe(2,0)(8,2)
	\rput(5,1){\scriptsize accumulator}
	\psline(5,2)(5,3)
	\pspolygon[fillstyle=hlines](4,3)(6,3)(7,4)(3,4)
	\pspolygon(1,5)(3,5)(4,6)(0,6)
	\psline(2,5)(2,4.5)(4,4.5)(4,4)
	\pspolygon(7,5)(9,5)(10,6)(6,6)
	\psline(8,5)(8,4.5)(6,4.5)(6,4)
	\psframe(0.25, 7)(1.75, 8.5)
	\rput(1,7.75){\scriptsize $-s$}
	\psline(1,7)(1,6)
	\psframe(2.25, 7)(3.75, 8.5)
	\rput(3,7.75){\scriptsize $s$}
	\psline(3,7)(3,6)
	\psframe(6.25,7)(7.75,8.5)
	\rput(7,7.75){\scriptsize $-1$}
	\psline(7,7)(7,6)
	\psline(9,6)(9,9)(1,9)(1,8.5)
	\psline(3,8.5)(3,9)
	\psline(7,8.5)(7,9)
	\psline(5,9)(5,10)
	\psline(5,0)(5,-1)
	\psline(7,2)(7,2.5)(12,2.5)(12,-1)
	\rput[t](5,-1){\scriptsize to memory}
	\rput[t](12,-1){\scriptsize from memory}
}
\psframe(2,10)(20,14)
\psframe[linestyle=dotted](-1,14.5)(23.5,24.5)
\rput[b]{90}(-1.2,19.5){\scriptsize input stage}
\psframe[linestyle=dotted](-1,-0.5)(23.5,9.5)
\rput[b]{90}(-1.2,4.5){\scriptsize output stage}
\rput(11, 12){\scriptsize $\F_{3^{97}}$-multiplier}
\psline(5,15)(5,14)
\psline(17,15)(17,14)
\end{pspicture}
\caption{The proposed structure block for implementing the formulas of
  Appendix~\ref{sec:appendix}}
\label{fig:proposedstructure}
\end{figure}

All of the $\F_{3^{97}}$-multiplications can be done in parallel. This
property allows designers to implement as many of these multipliers as
possible, according to their time-area constraints. On the other hand,
these multipliers are used for other computations such as point
addition and doubling on elliptic curves for pairing-based
cryptography. Reading and writing intermediate values into register
files in such applications is time-consuming. 
To solve this problem we propose a new multiplier which is shown in Figure~\ref{fig:proposedstructure}. 
The new multiplier consists of three pipeline stages, namely, input, multiplication, and output. 
During the time of each multiplication in $\F_{3^{97}}$, the input stage loads the coefficients $a_i$ and $b_i$ from memory for the next multiplication, and computes the linear combinations in (\ref{equ:formulasinput}) to compute $P_i$s.
In this time the output stage adds the last computed product $P_i$ to memory variables according to (\ref{equ:formulasoutput}). 
In this structure the hatched multiplexers can select either one of their
inputs or the sum of the inputs. In this way all possible multiples of
input polynomials can be selected and added to the accumulators.

\section{Conclusion}
\label{sec:conclusion}

In this paper we proposed a new structure for multiplication in
$\F_{3^{97}}$. This structure is based on digit-level LFSR multipliers, 
where the area of digit-multipliers are reduced using the Karatsuba
method. 
Another advantage of this approach is performing the modular reduction
during the multiplication. Our synthesis results showed the
performance improvement compared to other designs in the
literature. We have also presented new formulas for multiplication in
$\F_{3^{6 \cdot 97}}$ using only $15$ multiplications in
$\F_{3^{97}}$.
When the Karatsuba method is applied 18 multiplications are required.
Furthermore, we have introduced a feasible hardware structure for realizing our proposed formulas. Our formulas are for the case that $\F_{3^{6 \cdot 97}}$ is
constructed from $\F_{3^{2 \cdot 97}}$ using the irreducible
polynomial $z^3-z-1$. In case that the finite field is constructed using
$z^3-z+1$, the formulas require slight modifications.

\appendix

\section{Multiplication formulas for $\F_{3^{6 \cdot 97}}$}
\label{sec:appendix}

Let $\alpha, \beta \in \F_{3^{6 \cdot 97}}$ be given as:
\begin{align*}
\alpha = & a_0+a_1s+a_2r+a_3rs+a_4r^2+a_5r^2s,\\
\beta = & b_0+b_1s+b_2r+b_3rs+b_4r^2+b_5r^2s,\\
\end{align*}
where $a_0, \cdots, b_5 \in \F_{3^{97}}$ and $s \in \F_{3^{2 \cdot
    97}}$, $r \in \F_{3^{6 \cdot 97}}$ are roots of $y^2+1$ and
$z^3-z-1$, respectively.
Let their product $\gamma = \alpha \beta \in \F_{3^{6 \cdot 97}}$ be 
\begin{align*}
\gamma = c_0+c_1s+c_2r+c_3rs+c_4r^2+c_5r^2s.\\
\end{align*}
Then the coefficients $c_0 \cdots c_5 \in \F_{3^{97}}$ of the product
can be computed using the following formulas.

\begin{equation}
\begin{array}{ll}
P_0 = & (a_0+a_2+a_4)(b0+b2+b4)\\
P_1 = & (a_0+a_1+a_2+a_3+a_4+a_5)\\
& (b_0+b_1+b_2+b_3+b_4+b_5)\\
P_2 = &(a_1+a_3+a_5)(b_1+b_3+b_5)\\
P_3 = &(a_0+sa_2-a_4)(b_0+sb_2-b_4)\\
P_4 = &(a_0+a_1+sa_2+sa_3-a_4-a_5)\\
& (b_0+b_1+sb_2+sb_3-b_4-b_5)\\
P_5 = &(a_1+sa_3-a_5)(b_1+sb_3-b_5)\\
P_6 = & (a_0-a_2+a_4)(b_0-b_2+b_4)\\
P_7 = & (a_0+a_1-a_2-a_3+a_4+a_5)\\
& (b_0+b_1-b_2-b_3+b_4+b_5)\\
P_8 = & (a_1-a_3+a_5)(b_1-b_3+b_5)\\
P_9 = & (a_0-sa_2-a_4)(b_0-sb_2-b_4)\\
P_{10} = & (a_0+a_1-sa_2-sa_3-a_4-a_5)\\
& (b_0+b_1-sb_2-sb_3-b_4-b_5)\\
P_{11} = & (a_1-sa_3-a_5)(b_1-sb_3-b_5)\\
P_{12} = & a_4b_4\\
P_{13} = & (a_4+a_5)(b_4+b_5)\\
P_{14} = & a_5b_5\\
\end{array}
\label{equ:formulasinput}
\end{equation}
\newpage
\begin{equation}
\begin{array}{ll}
c_0 = & -P_0+P_2+(s+1)P_3-(s+1)P_5-\\
& (s-1) P_9+(s-1)P_{11}-P_{12}+P_{14}\\
c_1 = & P_0-P_1+P_2-(s+1)P_3+(s+1)P_4-\\
& (s+1)P_5+ (s-1)P_9-(s-1)P_{10}+\\
& (s-1)P_{11}-P_{12}-P_{13}+P_{14}\\
c_2 = & -P_{0}+P_{2}+P_{6}-P_{8}+P_{12}-P_{14}\\
c_3 = & P_{0}-P_{1}+P_{2}-P_{6}+P_{7}-P_{8}-P_{12}\\
& +P_{13}-P_{14}\\
c_4 = & P_{0}-P_{2}-P_{3}+P_{5}+P_{6}-P_{8}-P_{9}+P_{11}+\\
& P_{12}-P_{14}\\
c_5 = & P_{0}+P_{1}-P_{2}+P_{3}-P_{4}+P_{5}-P_{6}+P_{7}-\\
& P_{8}+P_{9}-P_{10}+P_{11}-P_{12}+P_{13}-P_{14}
\end{array}
\label{equ:formulasoutput}
\end{equation}


\small

\begin{thebibliography}{10}
\providecommand{\url}[1]{#1}
\csname url@rmstyle\endcsname
\providecommand{\newblock}{\relax}
\providecommand{\bibinfo}[2]{#2}
\providecommand\BIBentrySTDinterwordspacing{\spaceskip=0pt\relax}
\providecommand\BIBentryALTinterwordstretchfactor{4}
\providecommand\BIBentryALTinterwordspacing{\spaceskip=\fontdimen2\font plus
\BIBentryALTinterwordstretchfactor\fontdimen3\font minus
  \fontdimen4\font\relax}
\providecommand\BIBforeignlanguage[2]{{%
\expandafter\ifx\csname l@#1\endcsname\relax
\typeout{** WARNING: IEEEtran.bst: No hyphenation pattern has been}%
\typeout{** loaded for the language `#1'. Using the pattern for}%
\typeout{** the default language instead.}%
\else
\language=\csname l@#1\endcsname
\fi
#2}}

\bibitem{knu98}
D.~E. Knuth, \emph{The Art of Computer Programming, vol. 2, Seminumerical
  Algorithms}, 3rd~ed.\hskip 1em plus 0.5em minus 0.4em\relax Reading~MA:
  Addison-Wesley, 1998, first edition 1969.

\bibitem{gatger03}
J.~von~zur Gathen and J.~Gerhard, \emph{Modern Computer Algebra}, 2nd~ed.\hskip
  1em plus 0.5em minus 0.4em\relax Cambridge,~UK: Cambridge University Press,
  2003, first edition 1999.

\bibitem{dss00}
\BIBentryALTinterwordspacing
\emph{Digital Signature Standard (DSS)}, U.S. Department of Commerce / National
  Institute of Standards and Technology, January 2000, federal Information
  Processings Standards Publication 186-2. [Online]. Available:
  \url{http://csrc.nist.gov/publications/fips/fips186-2/fips186-2.pdf}
\BIBentrySTDinterwordspacing

\bibitem{baipaa98}
D.~V. Bailey and C.~Paar, ``Optimal extension fields for fast arithmetic in
  public-key algorithms,'' in \emph{Advances in Cryptology: Proceedings of
  CRYPTO~'98, {\rm Santa Barbara~CA}}, ser. Lecture Notes in Computer Science,
  H.~Krawczyk, Ed., no. 1462.\hskip 1em plus 0.5em minus 0.4em\relax
  Springer-Verlag, 1998, pp. 472--485.

\bibitem{avamih03}
R.~M. Avanzi and P.~Mih{\u{a}}ilescu, ``Generic efficient arithmetic algorithms
  for {{PAFF}}s (processor adequate finite fields) and related algebraic
  structures (extended abstract),'' in \emph{Selected Areas in Cryptography
  (SAC 2003)}.\hskip 1em plus 0.5em minus 0.4em\relax Springer-Verlag, 2003,
  pp. 320--334.

\bibitem{duulee03}
\BIBentryALTinterwordspacing
I.~Duursma and H.~Lee, ``Tate-pairing implementations for tripartite key
  agreement.'' [Online]. Available:
  \url{citeseer.ist.psu.edu/duursma03tatepairing.html}
\BIBentrySTDinterwordspacing

\bibitem{kermar05}
T.~Kerins, W.~P. Marnane, E.~M. Popovici, and P.~S. L.~M. Barreto, ``Efficient
  hardware for the tate pairing calculation in characteristic three,'' in
  \emph{Cryptographic Hardware and Embedded Systems, CHES2005}, ser. Lecture
  Notes in Computer Science, J.~R. Rao and B.~Sunar, Eds., vol. 3659.\hskip 1em
  plus 0.5em minus 0.4em\relax Springer-Verlag, 2005, pp. 412--426.

\bibitem{bergua03}
G.~Bertoni, J.~Guajardo, S.~Kumar, G.~Orlando, C.~Paar, and T.~Wollinger,
  ``Efficient {$GF(p^m)$} arithmetic architectures for cryptographic
  applications,'' in \emph{{Topics in cryptology, {CT-RSA 2003}: the
  Cryptographers' Track at the {RSA} Conference 2003, San Francisco, {CA},
  {USA}, April 13--17, 2003: Proceedings}}, ser. Lecture Notes in Computer
  Science, M.~Joye, Ed., vol. 2612.\hskip 1em plus 0.5em minus 0.4em\relax
  Springer-Verlag, 2003, pp. 158--175.

\bibitem{mce87}
R.~J. McEliece, \emph{{F}inite {F}ields for {C}omputer {S}cientists and
  {E}ngineers}.\hskip 1em plus 0.5em minus 0.4em\relax Kluwer Academic
  Publishers, 1987.

\bibitem{sho06}
\BIBentryALTinterwordspacing
J.~Shokrollahi, ``Efficient implementation of elliptic curve cryptography on
  fpgas,'' Ph.D. dissertation, Bonn University, Bonn, December 2006. [Online].
  Available:
  \url{http://hss.ulb.uni-bonn.de/diss\_online/math\_nat\_fak/2007/shokrollahi%
\_jamshid/ index.htm}
\BIBentrySTDinterwordspacing

\bibitem{grapag05}
\BIBentryALTinterwordspacing
R.~Granger, D.~Page, and M.~Stam, ``Hardware and software normal basis
  arithmetic for pairing-based cryptogaphy in characteristic three,''
  \emph{IEEE Transactions on Computers}, vol.~54, no.~7, pp. 852--860, 2005.
  [Online]. Available: \url{http://}
\BIBentrySTDinterwordspacing

\bibitem{gatsho06}
J.~von~zur Gathen and J.~Shokrollahi, ``{Fast arithmetic for polynomials over
  $\F_{2}$ in hardware},'' in \emph{IEEE Information Theory Workshop
  (2006)}.\hskip 1em plus 0.5em minus 0.4em\relax Punta del Este, Uruguay:
  IEEE, March 2006, pp. 107--111.

\bibitem{karofm63}
A.~Karatsuba and Y.~Ofman, ``Multiplication of multidigit numbers on
  automata,'' \emph{Soviet Physics--Doklady}, vol.~7, no.~7, pp. 595--596,
  January 1963, translated from Doklady Akademii Nauk SSSR, Vol.~145, No.~2,
  pp.~293--294, July, 1962.

\end{thebibliography}
\end{document}